\documentclass[aps,prb,twocolumn,tightenlines,nofootinbib]{revtex4}
\usepackage{bm}
\usepackage{graphics}
\usepackage{epsfig}
\usepackage{graphicx}
\usepackage{amsmath}
\usepackage{amsfonts}
\usepackage{amssymb}
\usepackage{bbold}
\usepackage{color}
\begin{document}

\title{Infinite cascades of phase transitions in the classical Ising chain}
\author{P. N. Timonin}
%\email{pntim@live.ru}
\affiliation{Physics Research Institute,
Southern Federal University, 194 Stachki ave.,  Rostov-on-Don, 344090 Russia}
\author{Gennady  Y. Chitov }
%\email{gchitov@laurentian.ca}
\affiliation{Department of Physics, Laurentian University, Sudbury, ON,
P3E 2C6 Canada}
\affiliation{Department of Physics,
McGill University, Montr\'{e}al, QC, H3A 2T8 Canada }

\date{\today}

\begin{abstract}
We report the new exact results on one of the best studied models in statistical physics: the classical
antiferromagnetic Ising chain in a magnetic field. We show that the model possesses an infinite cascade of
thermal phase transitions (also known as ``disorder lines" or geometric phase transitions). The phase transition is signalled by a change of
asymptotic behavior of the nonlocal string-string correlation functions when their monotonous decay becomes modulated by
incommensurate oscillations. The transitions occur for rarefied ($m$-periodic) strings with arbitrary odd $m$.
We propose a duality transformation which maps the Ising chain onto the $m$-leg Ising tube with nearest-neighbor
couplings along the legs and the plaquette four-spin interactions of adjacent legs. Then the $m$-string correlation
functions of the Ising chain are mapped onto the two-point spin-spin correlation functions along the legs of the $m$-leg tube.
We trace the origin of these cascades of  phase transitions to
the lines of the Lee-Yang zeros of the Ising chain in $m$-periodic complex magnetic field,
allowing us to relate these zeros to the observable
(and potentially measurable) quantities.
\end{abstract}

\maketitle

%
%
%%%%%%%%%%%%%%%%%%%%%%%%%%%%%%%%%%%%%%%%%%%%%%%%%%%%%%%%%%%%%%%%%%%%%%%%%%%%%%
\section{Introduction and Motivation. Disorder line in the classical Ising chain}\label{Intro}
%%%%%%%%%%%%%%%%%%%%%%%%%%%%%%%%%%%%%%%%%%%%%%%%%%%%%%%%%%%%%%%%%%%%%%%%%%%%%%
%
%
%%%%%%%%%%%%%%%%%%%%%%%%%%%%%%%%%%%%%%%%%%%%%%%%%%%%%%%%%%%%%%%%%%%%%%%%%%%%%%%%%%%%%%%%%%%%%%%%%%%%%
%%%%%%%%%%%%%%%%%%%%%%%%%%%%%%%%%%%%%%%%%%%%%%%%%%%%%%%%%%%%%%%%%%%%%%%%%%%%%%%%%%%%%%%%%%%%%%%%%%%%%
%%%%%%%%%%%%%%%%%%%%%%%%%%%%%%%%%%%%%%%%%%%%%%%%%%%%%%%%%%%%%%%%%%%%%%%%%%%%%%%%%%%%%%%%%%%%%%%%%%%%%
%%%%%%%%%%%%%%%%%%%%%%%%%%%%%%%%%%%%%%%%%%%%%%%%%%%%%%%%%%%%%%%%%%%%%%%%%%%%%%%%%%%%%%%%%%%%%%%%%%%%%

Yang and Lee \cite{LeeYang52} pioneered a rigorous direction of research in phase transitions relating them
to zeros of model's partition function. The original analysis of the ferromagnetic Ising model
created a huge impact and a vast literature extending results \cite{LeeYang52} on other models and/or non-equilibrium cases.
For a very limited list of references see, e.g.,  \cite{Fisher,Suzuki91,Bena05,Matveev08,Lebowitz16}.

There are however transitions which appear to be at odds with the Lee-Yang results. One of examples is the percolation transition
on the Kert\'{e}sz line $H_K(T)$ in the magnetic field--temperature $(H,T)$ plane of the 2D Ising model.  \cite{Kertesz89} Since the partition function of this model does not possess zeros of the complex field $H \in \mathbb{C}$ \cite{LeeYang52} such that $H \to H_K(T)$ in the thermodynamic limit, the word ``geometric" is used in the literature to emphasize an apparent distinction of such transitions from conventional ones. The very recent findings of infinite cascades of geometric
percolative transitions in several simple classical models \cite{UsPerc} made the problem even more challenging: How this infinite sequence
of transitions can be encoded in the analytical properties of the model's partition function? The extra problem with percolation
transitions in 2D models \cite{Kertesz89,UsPerc} is that it is very hard to deal analytically with nonlocal percolative order parameters,
so analysis is mainly restricted to numerical simulations.

To understand how the aforementioned geometric transitions can be framed into the standard Lee-Yang paradigm, we will take a fresh look at the probably most venerable textbook example.\cite{Baxter} We will study the 1D Ising model in presence of the magnetic field with the Hamiltonian
\begin{equation}
\label{H}
  \beta \mathcal{H} = K \sum_{n=1}^N \big(s(n) s(n+1)-1 \big) -h \sum_{n=1}^N s(n)~,
\end{equation}
where $s(n)= \pm 1$ are the Ising variables with periodic boundary conditions (PBC) $s(N+1)=s(1)$,
$\beta=1/ k_B T$ is the inverse temperature ($k_B=1$ is set in the following).
We consider the antiferromagentic nearest-neighbor exchange coupling $J>0$ ($K \equiv \beta J$). Since the
Hamiltonian is invariant under the simultaneous sign change of the magnetic field $H \leftrightarrow -H$ ($h \equiv \beta H$)
and spins $s(n) \leftrightarrow -s(n)$, it suffices to take $h>0$ in our formulas. In the definition \eqref{H} we subtracted
the constant term $NJ$ from the conventional Hamiltonian.  As known (see, e.g., \cite{Baxter}) this model is solved utilizing
the $2 \times 2$ transfer matrix. We choose it as $U_{ss'}=\exp [ K(1-ss')+hs ]$, or
\begin{equation}
\label{U}
  \hat{U}(h)= e^{h \hat{\sigma}_z} \big( \hat{\mathbb{1}} + e^{2K} \hat{\sigma}_x  \big)~,
\end{equation}
written in terms of the standard Pauli matrices $\hat{\sigma}_\sharp$. The eigenvalues of $\hat{U}$ are
\begin{equation}
\label{Lampm}
  \lambda_\pm (h)= \cosh h \pm R(h), ~~R(h) \equiv \sqrt{e^{4K}+ \sinh^2 h }~.
\end{equation}
The partition function of the model
\begin{equation}
\label{Z}
  Z(h)=\mathrm{ Tr} \hat{U}^N= \lambda_+^N(h)+  \lambda_-^N(h)~.
\end{equation}
is analytic and positive for any $h \in \mathbb{R}$ and $T>0$, even in the limit $N \to \infty$,
so the free energy is an analytic function as well.  Thus no phase transition is expected at nonzero temperature.
However Stephenson \cite{Stephenson70} found a phase transition on a so-called ``disorder line", which is an analogue of the
geometric transition discussed above.

\begin{figure}[h]
\centering{\includegraphics[width=7.0 cm] {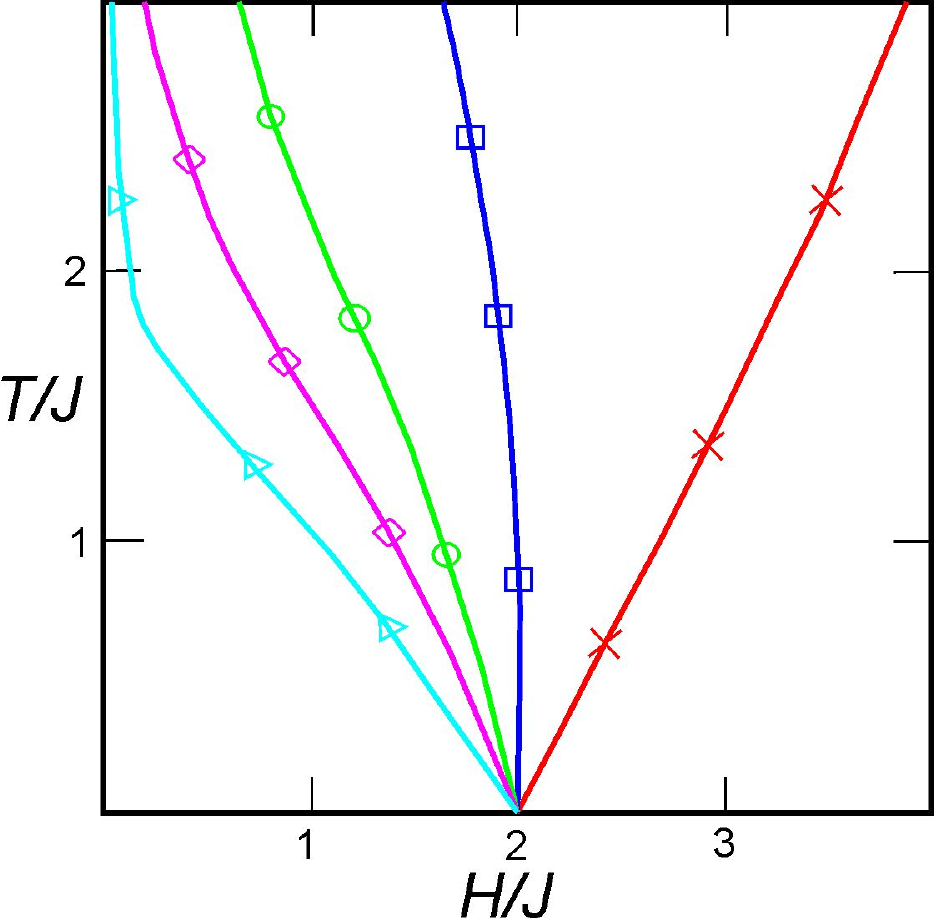}}
 \caption{(Color online) Phase diagram of the Ising model with several disorder lines $T_{c,m}(H)$ explicitly plotted.
 The red curve with crosses located at $H \geq 2 J$ is the original disorder line $T_c(H)$ ($m=1$) found by Stephenson.
 The newly found lines of phase transitions $T_{c,m}(H)$ ($m\geq 3$) filling the region $H \leq 2 J$ are monotonously
 decreasing functions of both $m$ and $H$.  Shown in the figure: $\times$  (m=1): $\square$  (m=3); $\bigcirc$  (m=5);
 $\Diamond$ (m=7); $\triangle$  (m=11).} \label{CritLines}
\end{figure}

To understand this result let us first introduce the string operators
\begin{equation}
\label{String}
  \sigma(n)  \equiv \prod_{k=1}^{n} s(k) = (-i)^n
  \exp \Big[\frac{i\pi}{2}\sum_{k=1}^{n} s(k) \Big]~,
\end{equation}
which are also Ising variables $\sigma(n) = \pm 1$, and they provide a dual representation
\cite{Dobson69} of the Hamiltonian \eqref{H}:
\begin{equation}
\label{Hnnn}
  \beta \mathcal{H} = K \sum_{n=1}^N \big(\sigma(n) \sigma(n+2)-1 \big) -h \sum_{n=1}^N \sigma(n) \sigma(n+1)~,
\end{equation}
known as the Anisotropic Next-Nearest Neighbor Ising (ANNNI) chain. \cite{Liebmann86} The nonlocal string-string correlation function of the Ising chain \eqref{H} is a conventional two-point correlation function of the dual ANNNI model \eqref{Hnnn}:
\begin{equation}
\label{Gr}
  G(r) \equiv \langle \sigma(l) \sigma(l+r)  \rangle  = \Big\langle \prod_{k=l+1}^{l+ r} s(k)  \Big\rangle  ~.
\end{equation}
Using the approaches of \cite{Marsh66,Dobson69}, Stephenson \cite{Stephenson70} found the phase transition in the chain along the disorder line $H_c(T)$ shown in Fig.~\ref{CritLines}.  It is determined from equation:
\begin{equation}
\label{hc}
 h_c(K)=\mathrm{arccosh} ~e^{2K} = \log \big(e^{2K} +\sqrt{e^{4K}-1} \big) ~.
\end{equation}

The transition consists in changing asymptotic behavior of the correlation function $G(r)$ at large $r \gg 1$.  For  $H >H_c$ the correlations
decay exponentially $G(r) \sim e^{- \kappa r}$, while for $H < H_c$ this monotonous decay is modified by superimposed oscillations as
%\begin{equation}
%\label{Gos}
  $G(r) \sim e^{- \kappa r} \cos(q r +\delta)$
%\end{equation}
with an incommensurate wave vector $q$ given by
\begin{equation}
\label{q}
 q = \arcsin \sqrt{\frac{e^{4K}- \cosh ^2 h}{e^{4K}-1}  }  ~.
\end{equation}
Similar disorder lines exist in other Ising models \cite{Liebmann86},
in the quantum $XY$ chain in transverse field \cite{Barouch71}, or in the classical $\mathcal{O}(n)$ spin models \cite{Nussinov12}.

%
%
%%%%%%%%%%%%%%%%%%%%%%%%%%%%%%%%%%%%%%%%%%%%%%%%%%%%%%%%%%%%%%%%%%%%%%%%%%%%%%
\section{Disorder line and complex magnetic field}\label{DLF}
%%%%%%%%%%%%%%%%%%%%%%%%%%%%%%%%%%%%%%%%%%%%%%%%%%%%%%%%%%%%%%%%%%%%%%%%%%%%%%
%
%
%%%%%%%%%%%%%%%%%%%%%%%%%%%%%%%%%%%%%%%%%%%%%%%%%%%%%%%%%%%%%%%%%%%%%%%%%%%%%%%%%%%%%%%%%%%%%%%%%%%%%
%%%%%%%%%%%%%%%%%%%%%%%%%%%%%%%%%%%%%%%%%%%%%%%%%%%%%%%%%%%%%%%%%%%%%%%%%%%%%%%%%%%%%%%%%%%%%%%%%%%%%
%%%%%%%%%%%%%%%%%%%%%%%%%%%%%%%%%%%%%%%%%%%%%%%%%%%%%%%%%%%%%%%%%%%%%%%%%%%%%%%%%%%%%%%%%%%%%%%%%%%%%
%%%%%%%%%%%%%%%%%%%%%%%%%%%%%%%%%%%%%%%%%%%%%%%%%%%%%%%%%%%%%%%%%%%%%%%%%%%%%%%%%%%%%%%%%%%%%%%%%%%%%

To understand how the occurrence of this thermal phase transition found by Stephenson is compatible with the Lee-Yang theorem \cite{LeeYang52} and analytic $Z(h)>0$  given by equation \eqref{Z}, the following simple observation is crucial: The evaluation of $G(r)$ as an average of the string \eqref{String} with the Hamiltonian \eqref{H} for $n \sim N$ and in the thermodynamic limit $N \to \infty$ amounts to dealing with \eqref{H} in presence of the complex magnetic field $\mathfrak{h} = h +i \pi/2$. Thus, to diagnose the phase transition probed by nonlocal string correlation functions, one needs to study the analytical properties of $Z(h +i \pi/2)$.
An equivalent conclusion can be drawn if one chooses the dual framework.  Indeed, by using the identity
%\begin{equation}
%\label{Pair}
 $ \sigma(n)  \sigma(m)  =-i
  \exp \Big[\frac{i\pi}{2} \sigma(n)  \sigma(m)   \Big]$,
%\end{equation}
one can write
\begin{equation}
\label{CorrString}
  \langle \sigma(1) \sigma(n) \rangle = (-i)^n
 \Big\langle  \exp \Big[\frac{i\pi}{2}\sum_{k=1}^{n-1} \sigma(k) \sigma(k+1) \Big]  \Big\rangle~.
\end{equation}
Thus transition probed by the local spin-spin correlation function \eqref{CorrString} is due to the imaginary part $i \pi/2$ added to the nearest-neighbor exchange coupling. For the dual Hamiltoinian \eqref{Hnnn} the problem reduced again to evaluation of $Z(h +i \pi/2)$.

To formalize these observations we write the correlation function
\begin{equation}
\label{GnTrans}
  G(n) =\mathrm{ Tr} \big\{ (\hat{\sigma}_z \hat{U} )^n  \hat{U}^{N-n} \big\} /Z(h)
\end{equation}
via the transfer matrix with the complex magnetic field:
\begin{equation}
\label{V}
\hat{\sigma}_z \hat{U}(h) =-i  e^{ i \frac{\pi}{2}  \hat{\sigma}_z } \hat{U}(h)=
-i   \hat{U} \Big( h+ i \frac{\pi}{2} \Big)~.
\end{equation}
The Stephenson result on the disorder line of phase transitions is readily recovered  by setting $n = N \to \infty$ in the above formulas. The oscillations in $G(n) \propto Tr \{(-i \hat{U}(h+i \pi/2) )^n \} $ are due to appearance of the imaginary parts
in the eigenvalues $-i \lambda_\pm(h+i \pi/2)$ of the real matrix $\hat{\sigma}_z \hat{U}(h) = -i  \hat{U}(h+ i \pi/2)$
at $h<h_c(K)$  (cf. Eq.~\eqref{hc}.)

To relate this transition to the Lee-Yang zeros \cite{LeeYang52}, one can note that $Z(h+i \pi/2)$ is a polynomial over fugacity
$e^{2h}$, so it has zeros at some complex field $\mathfrak{h} \in \mathbb{C}$. In thermodynamic limit the locus of zeros is determined by
the conditions: \cite{Fisher,Suzuki91}
\begin{equation}
\label{YLcond}
  |\lambda_+(\mathfrak{h})|= |\lambda_-(\mathfrak{h})|,~~ \mathrm{ Im} \log \frac{\lambda_+(\mathfrak{h})}{\lambda_-(\mathfrak{h})} \neq 0~.
\end{equation}
It follows from  \eqref{Z} and  \eqref{Lampm} that in the thermodynamic limit the Lee-Yang zeros condense along the straight lines in the complex plane:
%\begin{equation}
%\label{ZeroLine}
   $\mathrm{Im} \mathfrak{h} =\pi/2 \pmod{\pi}, ~~|\mathrm{Re} \mathfrak{h}| < h_c(K),$
%\end{equation}
where the last inequality coincides with the Stephenson conditions for oscillations $ h < h_c(K)$.
Thus these oscillations relate the Lee-Yang zeros in complex fields to the observable (and potentially measurable) quantities.
Interestingly, results on this phase transition have been known for several decades by now \cite{Stephenson70}, but only recently \cite{Abanov13}, when they were re-derived under a different guise, it was also suggested that the oscillating asymptote of $G(n)$ may be somehow related to the Yang-Lee zeros.  To the best of our knowledge the rigorous relation between the transition on the disorder line and the Lee-Yang zeros is reported here for the first time.

%
%
%%%%%%%%%%%%%%%%%%%%%%%%%%%%%%%%%%%%%%%%%%%%%%%%%%%%%%%%%%%%%%%%%%%%%%%%%%%%%%
\section{Rarefied strings and infinite cascades of phase transitions}\label{RarStr}
%%%%%%%%%%%%%%%%%%%%%%%%%%%%%%%%%%%%%%%%%%%%%%%%%%%%%%%%%%%%%%%%%%%%%%%%%%%%%%
%
%
%%%%%%%%%%%%%%%%%%%%%%%%%%%%%%%%%%%%%%%%%%%%%%%%%%%%%%%%%%%%%%%%%%%%%%%%%%%%%%%%%%%%%%%%%%%%%%%%%%%%%
%%%%%%%%%%%%%%%%%%%%%%%%%%%%%%%%%%%%%%%%%%%%%%%%%%%%%%%%%%%%%%%%%%%%%%%%%%%%%%%%%%%%%%%%%%%%%%%%%%%%%
%%%%%%%%%%%%%%%%%%%%%%%%%%%%%%%%%%%%%%%%%%%%%%%%%%%%%%%%%%%%%%%%%%%%%%%%%%%%%%%%%%%%%%%%%%%%%%%%%%%%%
%%%%%%%%%%%%%%%%%%%%%%%%%%%%%%%%%%%%%%%%%%%%%%%%%%%%%%%%%%%%%%%%%%%%%%%%%%%%%%%%%%%%%%%%%%%%%%%%%%%%%

\begin{figure}[h]
\centering{\includegraphics[width=8.0 cm] {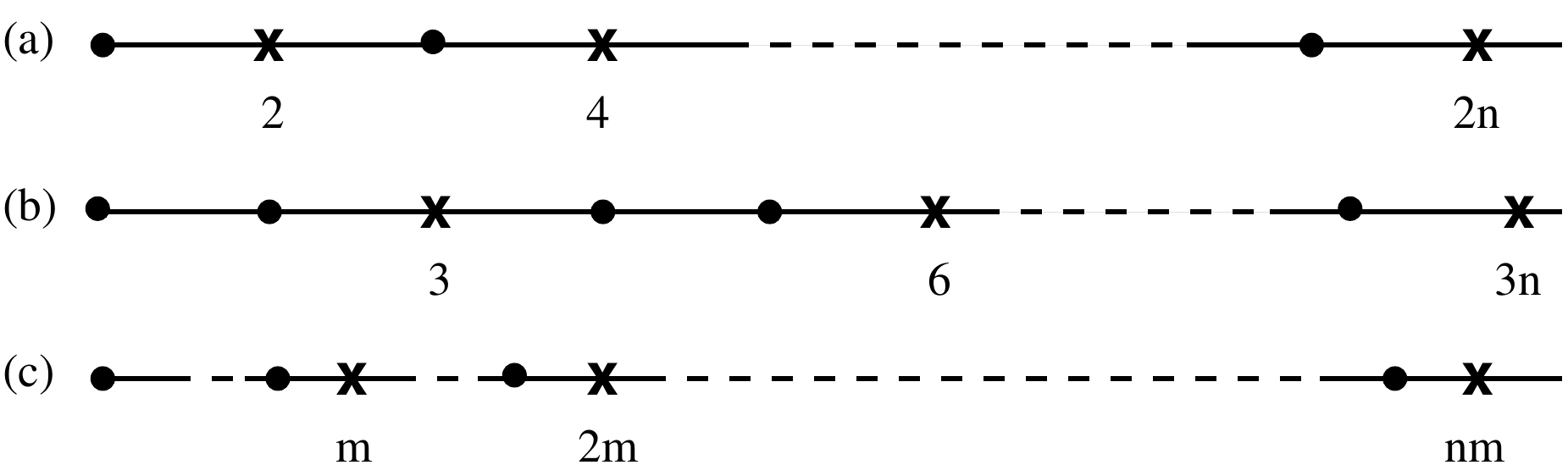}}
 \caption{Three examples of the rarefied strings: (a) $\sigma(n|2)$; (b) $\sigma(n|3)$; (c) $\sigma(n|m)$
 (not in scale). Crosses indicate spins $s(mk)$, $1 \leq k \leq n$ constituting the string. } \label{m-strings}
\end{figure}

We define rarefied strings made out of $n$ spins as
 \begin{equation}
\label{RarStr}
  \sigma(n|m) \equiv \prod_{k=1}^{n} s(mk)~.
\end{equation}
These operators are also Ising variables and generalize the simple string $\sigma(n|1) \equiv \sigma(n)$ defined by \eqref{String}.
Contrary to the latter
which includes all $n$ spins of the chain from its left end, the string $\sigma(n|m)$ includes $n$ spins $s(mk)$ residing on the sites
$m, 2m, ..., nm$ of the chain separated by $(m-1)$ spacings. The right end of this string is the $nm$-th site. See Fig.~\ref{m-strings}.

The new fundamental result we prove below is that the Ising model \eqref{H} possesses an infinite cascade of nonlocal phase transitions
on the multitude of disorder lines. Similarly to the known case \cite{Stephenson70}, the phase transition on each disorder line is
manifested by appearance of incommensurate oscillations in the corresponding correlation function of rarefied strings
\begin{equation}
\label{Gnm}
  G(n|m) \equiv \langle \sigma(l|m) \sigma(l+n|m)  \rangle  = \Big\langle \prod_{k=l+1}^{l+ n} s(mk)  \Big\rangle  ~.
\end{equation}
Note that this function  probes the correlations between the rarefied strings whose right ends are separated by $n$ incorporated spins and
by the distance of $r=nm$ spacings. These new critical points we found are also identified with loci of the Lee-Yang zeros of the partition function of the model \eqref{H} with the complex $m$-periodic magnetic field.

Let us write the correlation function \eqref{Gnm} as
\begin{equation}
\label{GnmUV}
  G(n|m) = \mathrm{Tr} \big\{ (\hat{\sigma}_z \hat{U}^m )^n  \hat{U}^{N-nm} \big\} /Z(h)~.
\end{equation}
Similarly to the case of homogeneous strings, the new phase transitions on the multiple disorder lines signaled by
oscillations in $G(n|m)$ correspond to appearance of the imaginary parts in
the eigenvalues  of $\hat V_m \equiv \hat{\sigma}_z \hat{U}^m$ which are 
\begin{eqnarray}
\label{muepsM}
\mu_m^\pm &=& \Delta^m \Big( \gamma_m \pm \sqrt{\gamma_m^2 -1} \Big) \nonumber \\
 &=& \Delta^m
\left\{
\begin{array}{lr}
\exp (\pm  \mathrm{arccosh}  \gamma_m ),
& \gamma_m>1~; \\
\exp (\pm i   \arccos  \gamma_m ),
& \gamma_m<1~.
\end{array}
\right.
\end{eqnarray}
where  $\Delta \equiv \sqrt{e^{4K}-1}$; $\gamma_m  \equiv \tanh h \, \tanh \zeta \, \cosh m \zeta$, and
\begin{equation}
\label{Zeta}
  \zeta \equiv  \mathrm{arcsinh} \Big( \frac{\cosh h}{\sqrt{e^{4 K}-1} } \Big)~.
\end{equation}
This is most readily seen by taking a particular choice $N=nm$ in \eqref{GnmUV} yielding $G(n|m) \propto  \mathrm{Tr} (\hat V_m)^n$. As we have shown in Appendix \ref{AppA}, the complex eigenvalues
$\mu_m^\pm$ (i.e. $\gamma_m<1$) are possible only for odd $m=2l+1,~ l \in \mathbb{Z}^+$,
which we will always assume, unless specified otherwise. Qualitatively, the occurrence of phase transitions (or complex
$\mu_m^\pm$) for odd $m$ only can be understood as a result of extra frustration due to coupling of two antiferromagnetic
sublattices (say $ABABAB...$) by the $m$-string operator. On the contrary, the even $m$-string operator includes spins from one sublattice only ($A$ or $B$).

From the above equations and the condition $\gamma_m=1$ one can find  (for the details, see Appendix \ref{AppA}) the equation for the critical (disorder) line
$T_{c,m}(H)$ of phase transition:
\begin{equation}
\label{Tcm}
  \sinh h \,\sinh m \zeta = e^{2K}~.
\end{equation}
At $T< T_{c,m}(H)$ (or, equivalently at $H>H_{c,m}(T)$) the correlation function $G(n|m)$ whose exact formula \eqref{GnmAbove} is calculated in Appendix \ref{AppB}, decays exponentially with the distance $r=nm \gg 1$ as
$G(n|m) \sim e^{- \kappa_m r}$.  At weaker fields $H<H_{c,m}(T)$ (or at
$T >T_{c,m}(H)$) we infer from the exact result \eqref{GnmBelow} for the correlation function $G(n|m)$ that  the exponent is modulated by oscillations as  $G(n|m) \sim e^{- \kappa_m r} \cos(q_m r +\delta_m)$.
The analytical expressions for the inverse correlation length $\kappa_m$ and the wave vector of oscillations $q_m$ are given by equations
\eqref{kappaAb}, \eqref{kappaBe} and \eqref{qm} in Appendix \ref{AppB}.

The phase diagram of the Ising chain with several disorder lines $T_{c,m}(H)$ numerically calculated from Eq.~\eqref{Tcm} for different $m$ is shown in Fig.~\ref{CritLines}. The number of critical lines of phase transitions filling up the region $H \leq 2J$ is infinite corresponding to any positive odd $m$. All those lines sprung from the same zero-temperature critical point $H=2J$ which separates the antiferromagnetic ($H<2J$)
and ferromagnetic ($H>2J$) ground states. The newly found lines of phase transitions in the correlation functions of rarefied strings $T_{c,m}(H)$ ($m\geq 3$) decrease monotonously with both $m$ and $H$. This implies that the ``truly disordered" phase with pure exponential decay of all string correlation functions $G(n|m)$ can be reached at low temperatures
$T< T_{c,1}$ only if $H>2J$. The phase diagram at $H< 2J$ consists only out of oscillating phases, and the number of simultaneously oscillating
$G(n|k)$ increases on the $(T,H)$-plane with decreasing $T$ or $H$. Each region between two adjacent critical lines
$T_{c,m+2}<T< T_{c,m}$ where $m=2l+1$, accommodates $l+1$ oscillating functions $G(n|k)$ with $k \leq m$. This infinite sequence of the critical lines resembles the devil's staircase of the commensurate-incommensurate transitions known for the 3D Ising models with competing couplings. \cite{Liebmann86,GC2004} 

\begin{figure}[h]
\centering{\includegraphics[width=8.0 cm]{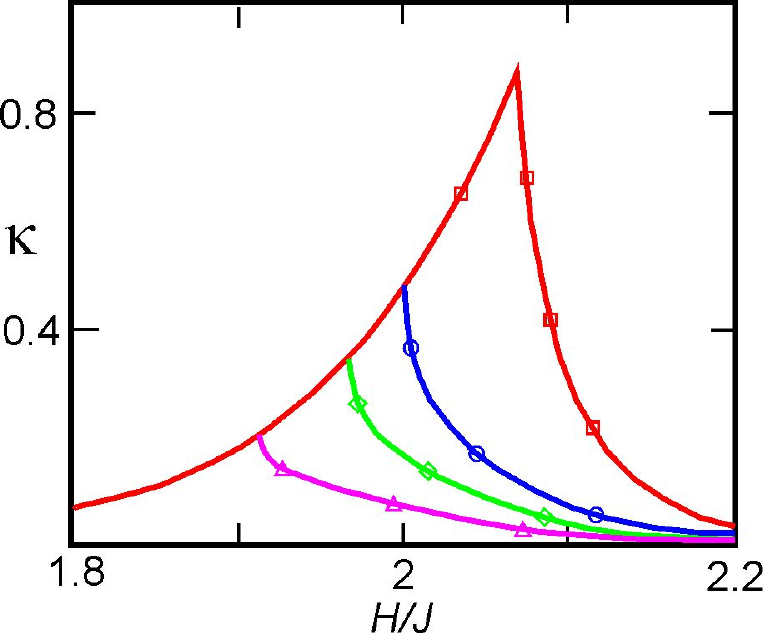}}
 \caption{(Color online) Inverse correlation lengths $\kappa_m$ as functions of the magnetic field at $T/J=0.1$ and for
 different values of $m$: $\square$  (m=1); $\bigcirc$  (m=3); $\Diamond$ (m=5); $\triangle$  (m=11).} \label{kappaFig}
\end{figure}

\begin{figure}[h]
\centering{\includegraphics[width=8.0 cm]{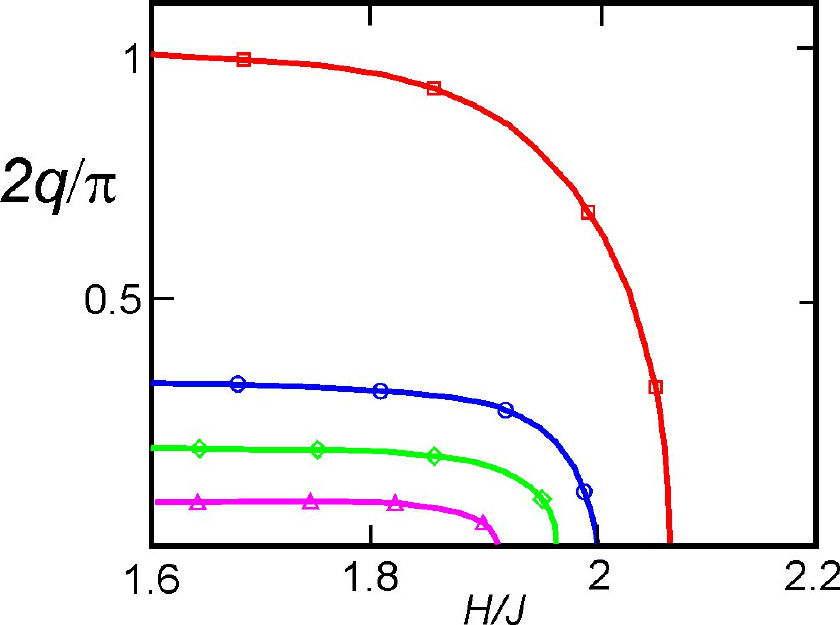}}
 \caption{(Color online) Wave vector of oscillations $q_m$ as a function of the magnetic field at $T/J=0.1$ and for
 different values of $m$: $\square$  (m=1); $\bigcirc$  (m=3); $\Diamond$ (m=5); $\triangle$  (m=11).} \label{qFig}
\end{figure}

However the lack of local or string order parameters at $T>0$ and finite correlation length at $T_{c,m}$ indicate at a very particular nature of the phase transition on the disorder lines.  The correlation length manifests a weaker singularity than expected for a continuous phase transition:
namely it has a cusp at the critical point, as one can see in Fig.~\ref{kappaFig}.
(Note that the name itself of the ``disorder line" \cite{Stephenson70} is due to the local minimum of the correlation length at the critical point.)  The wave vector of oscillations $q_m$
evolves smoothly from zero at the critical point towards the commensurate limit $q_m = \pi/2m$ at $H \to 0$, see Fig.~\ref{qFig}.
Such behavior is quite similar to that of the floating phase occurring in some frustrated 2D Ising models. \cite{BakVillain,UsIsing}

%
%
%

%
%
%%%%%%%%%%%%%%%%%%%%%%%%%%%%%%%%%%%%%%%%%%%%%%%%%%%%%%%%%%%%%%%%%%%%%%%%%%%%%%
\section{Dual $m$-leg spin tubes}\label{Mtubes}
%%%%%%%%%%%%%%%%%%%%%%%%%%%%%%%%%%%%%%%%%%%%%%%%%%%%%%%%%%%%%%%%%%%%%%%%%%%%%%
%
%
%%%%%%%%%%%%%%%%%%%%%%%%%%%%%%%%%%%%%%%%%%%%%%%%%%%%%%%%%%%%%%%%%%%%%%%%%%%%%%%%%%%%%%%%%%%%%%%%%%%%%
%%%%%%%%%%%%%%%%%%%%%%%%%%%%%%%%%%%%%%%%%%%%%%%%%%%%%%%%%%%%%%%%%%%%%%%%%%%%%%%%%%%%%%%%%%%%%%%%%%%%%
%%%%%%%%%%%%%%%%%%%%%%%%%%%%%%%%%%%%%%%%%%%%%%%%%%%%%%%%%%%%%%%%%%%%%%%%%%%%%%%%%%%%%%%%%%%%%%%%%%%%%
%%%%%%%%%%%%%%%%%%%%%%%%%%%%%%%%%%%%%%%%%%%%%%%%%%%%%%%%%%%%%%%%%%%%%%%%%%%%%%%%%%%%%%%%%%%%%%%%%%%%%

\begin{figure}[h]
\centering{\includegraphics[width=7.0 cm]{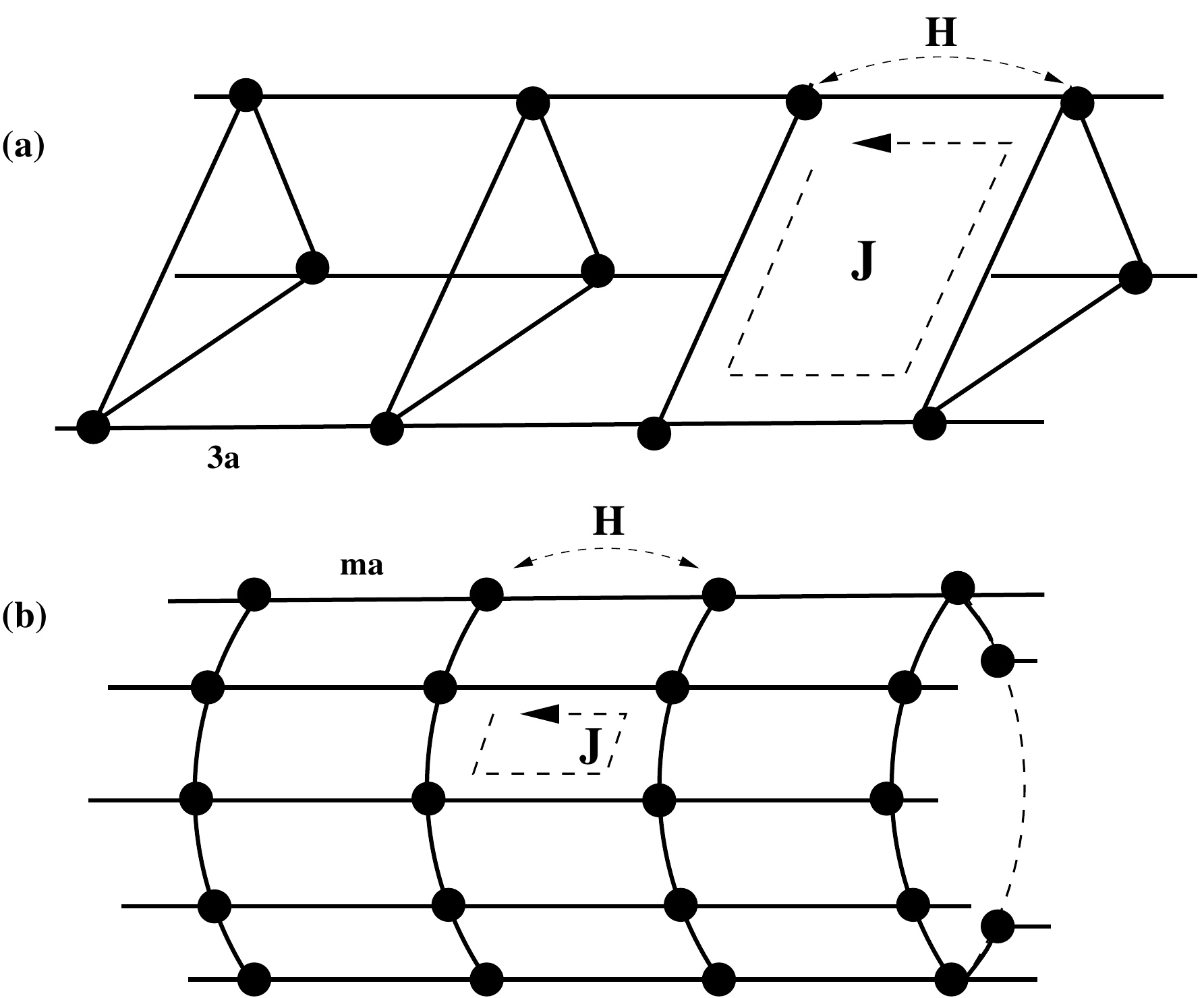}}
 \caption{ $m$-leg spin tubes dual to the Ising model \eqref{H}. (a) $m=3$,  the nearest-neighbor coupling along the leg is $H$;
  dashed loop on a plaquette indicates the four-spin coupling $J$.
  (b) fragment of the $m$-tube, similar notations. For consistency in the spatial dependence of $G(n|m)$
   the distance between the rungs of the tubes is set to $ma$, where $a$ is the spacing of the original Ising chain.} \label{tubeFig}
\end{figure}

It is interesting to push further the  (spin-string) duality between the Ising \eqref{H} and the ANNNI \eqref{Hnnn} chains to detect the cascades of transitions in terms of usual spin-spin correlation functions.  To proceed let us define more species of rarefied strings as
\begin{equation}
  \label{RarStrL}
  \sigma_\alpha(n|m) \equiv \prod_{k=1}^{n} s(mk-\alpha+1)~; ~~\alpha=1,...,m,
\end{equation}
which are just the strings \eqref{RarStr} (see Fig.~\ref{m-strings}) shifted by $\alpha-1$ spacings to the left.
Since  $\sigma_\alpha(n|m) \sigma_\alpha(n-1|m) =  s(mn-\alpha+1)$, the Ising model \eqref{H} can be mapped onto
the Hamiltonian of $m$-leg tube. With the choice $N =m N_L$ and the PBC along the chain, we imply periodicity in the $\alpha$-space:
$ \sigma_1(n|m) =\sigma_{m+1} (n|m)$. Simplifying notations further as $\tilde s_\alpha(n) \equiv \sigma_\alpha(n|m)$,
we get the dual representation of the Hamiltonian \eqref{H}:
\begin{eqnarray}
\label{Htube}
 \beta \mathcal{H} &=&  \sum_{n=1}^{N_L} \sum_{\alpha =1}^m   \Big\{ -h \tilde s _\alpha(n) \tilde s_\alpha(n+1) -1 \nonumber \\
  &+&K \tilde s_\alpha(n) \tilde s_\alpha(n+1) \tilde s_{\alpha+1} (n) \tilde s_{\alpha+1} (n+1) \Big\}~.
\end{eqnarray}
Here $\alpha$ is leg's number in the tube, $n$ numbers the spin in a leg, and $N_L$ is the number of spins in each leg.
The $m$-leg tube (see Fig.~\ref{tubeFig}) has the nearest-neighbor in-leg coupling $H$ and the plaquette coupling $J$ between four spins residing on the vertices of each facet of the tube. This chain-tube duality maps the nonlocal phase transition probed by $m$-periodic strings of the Ising chain \eqref{H} onto the disorder line of the spin-spin in-leg correlation function of the $m$-leg tube \eqref{Htube}:
\begin{eqnarray}
  \label{GnmTube}
  G(n|m) &=& \langle \sigma_\alpha(l|m) \sigma_\alpha(l+n|m) \rangle  \nonumber \\
  &\equiv&  \langle  \tilde s_\alpha(l) \tilde s_{\alpha} (l+n) \rangle~,~~ \forall~\alpha=1,...,m.
\end{eqnarray}
Contrary to some other examples of the spin chains or ladders (for more details, review and references, see, e,g, \cite{GT2017}), this
duality does not help to reveal the local or string Landau-type order parameters, since the correlation functions $G(n|m)$ ($\forall~m=2l+1$) decay on both sides from the critical points on the disorder lines $T_{c,m}(H)$.

%
%
%%%%%%%%%%%%%%%%%%%%%%%%%%%%%%%%%%%%%%%%%%%%%%%%%%%%%%%%%%%%%%%%%%%%%%%%%%%%%%
\section{Cascades of transitions and Lee-Yang zeros}\label{CascLYZ}
%%%%%%%%%%%%%%%%%%%%%%%%%%%%%%%%%%%%%%%%%%%%%%%%%%%%%%%%%%%%%%%%%%%%%%%%%%%%%%
%
%
%%%%%%%%%%%%%%%%%%%%%%%%%%%%%%%%%%%%%%%%%%%%%%%%%%%%%%%%%%%%%%%%%%%%%%%%%%%%%%%%%%%%%%%%%%%%%%%%%%%%%
%%%%%%%%%%%%%%%%%%%%%%%%%%%%%%%%%%%%%%%%%%%%%%%%%%%%%%%%%%%%%%%%%%%%%%%%%%%%%%%%%%%%%%%%%%%%%%%%%%%%%
%%%%%%%%%%%%%%%%%%%%%%%%%%%%%%%%%%%%%%%%%%%%%%%%%%%%%%%%%%%%%%%%%%%%%%%%%%%%%%%%%%%%%%%%%%%%%%%%%%%%%
%%%%%%%%%%%%%%%%%%%%%%%%%%%%%%%%%%%%%%%%%%%%%%%%%%%%%%%%%%%%%%%%%%%%%%%%%%%%%%%%%%%%%%%%%%%%%%%%%%%%%

\begin{figure}[h]
\centering{\includegraphics[width=8.5 cm] {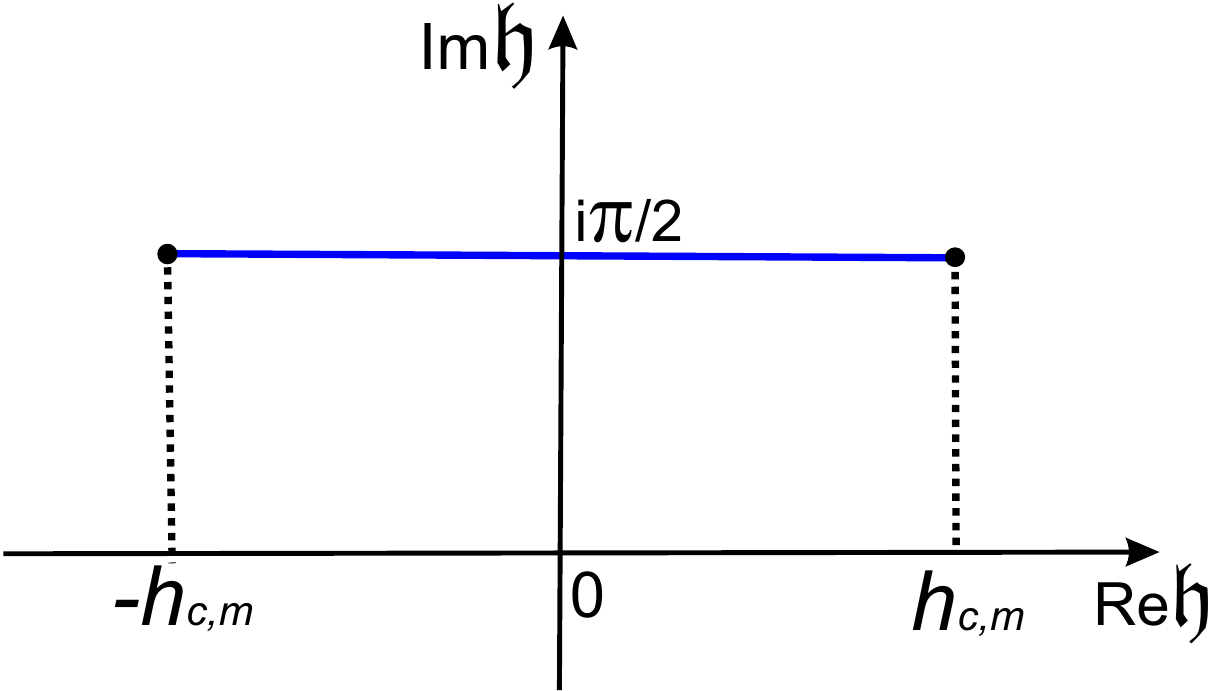}}
 \caption{(Color online) Loci of zeroes of the partition function on the plane of the homogeneous complex magnetic field
 $\mathfrak{h}$. Note that the line of the Lee-Yang zeros is periodic with respect to the shift of
  $\mathrm{Im} \mathfrak{h}$ by $\pi m $,  $m\in \mathbb{Z}$, and only one (blue bold) line with $\mathrm{Im} \mathfrak{h} = \pi/2$
  is plotted. On the real axis the zeros are bounded by the condition $ |\mathrm{Re} \mathfrak{h}| \leq h_c(K) 
  \equiv h_{c,1}(K)$. The figure
  applies also for the component $\mathfrak{h}$ of the m-periodic complex field $\mathfrak{h}_m$  defined by Eq.~\eqref{Hm}.
  In the latter case the bounds of the abscissa of the Lee-Yang line(s) are $ |\mathrm{Re} \mathfrak{h}| =h  \leq h_{c,m}(K)$.
  }\label{H_plane}
\end{figure}

We can now extend our previous result for the homogenous strings
and prove that all oscillating phases we found correspond to the Lee-Yang zeros of the Ising chain in various $m$-periodic
complex magnetic fields. Indeed, averaging of the string operator $\sigma(n|m)$ (cf. Fig.~\ref{m-strings}) with $N_L=N/m$ nodes
is equivalent to evaluation of the partition function of the Ising model with additional imaginary field $i \pi/2$ turned on at each crossed site shown in Fig.~\ref{m-strings}. According to Eq.~\eqref{GnmUV}
$G(N_L|m) \propto  \mathrm{Tr}\hat{V}_m^{N_L}$, where
\begin{eqnarray}
\label{VmCompl}
  \hat{V}_m(h)  &\equiv&   \hat{\sigma}_z \hat{U}^m(h)  = -i \hat{U}(\mathfrak{h}=h+i \pi/2 )\hat{U}^{m-1}(h)  \nonumber \\
   &\equiv &   -i \hat{\mathcal{U}}_m(\mathfrak{h}_m) \big|_{\mathfrak{h}=h +i \pi/2}~,
\end{eqnarray}
i.e., $\hat{V}_m$  is (up to a factor $-i$) the transfer matrix  $\hat{\mathcal{U}}_m$ of the Ising chain in the complex-valued $m$-periodic magnetic field
\begin{equation}
\mathfrak{h}_m(k)=
\left\{
\begin{array}{lr}
\mathfrak{h} \in \mathbb{C},
& k=m l~ \& ~l\in \mathbb{N} ; \\
h
& \mathrm{elswhere}~,
\end{array}
\right. \label{Hm}
\end{equation}
evaluated at a particular value $\mathfrak{h}=h +i \pi/2$. Thus we can identify the locus of the Lee-Yang zeros of the Ising model
with the periodic field $\mathfrak{h}_m$ and the region in the parametric space where the oscillating phase is located. Since
\begin{equation}
\label{LYz}
  (-i)^{N_L} Z(\mathfrak{h}_m)\big|_{\mathfrak{h}=h +i \pi/2}=(\mu_m^+)^{N_L}+(\mu_m^-)^{N_L}~,
\end{equation}
zeros of the partition function appear when the eigenvalues $\mu_m^\pm$ of $\hat{V}_m$ become complex, and
since $\hat{V}_m \in \mathbb{R}$, they are complex conjugate: $(\mu_m^+)^{\ast}=\mu_m^-$. With the results given in Appendix \ref{AppA} for the case when the auxiliary parameter \eqref{eps} $\gamma_m<1$,
we infer that the partition function $Z(\mathfrak{h}_m)\big|_{\mathfrak{h}=h +i \pi/2} \propto \cos Nq_m$ is a rapidly oscillating function with $\mathcal{O} (N)$ zeros in the complex plane  $\mathfrak{h}$. In the thermodynamic limit the locus of zeros condenses into a line shown in Fig.~\ref{H_plane}:
\begin{equation}
\label{ZeroLineM}
   \mathrm{Im} \mathfrak{h} =\pi/2 \pmod{\pi}, ~~|\mathrm{Re} \mathfrak{h}|=h \leq  h_{c,m}(K)~,
\end{equation}
with the critical field  $h_{c,m}(K)$ determined by Eq.~\eqref{Tcm}.

We need to emphasize that although we evaluate the thermodynamic quantities using a complex magnetic field, \textit{it does not make them identical} to their counterparts for the model where such field is turned on. For example, the $m$-string correlation function
$G(N_L|m) = (-i)^{N_L} Z(\mathfrak{h}_m)/Z(h)$ for $\mathfrak{h}=h +i \pi/2$, i.e., the average is weighted with the partition function $Z(h)$ for the Hamiltonian \eqref{H} with the real magnetic field $h$. In Appendix \ref{AppC} we present several results for the Ising chain
with $m$-periodic complex field  \eqref{Hm} with $\mathfrak{h}=h +i \pi/2$. The thermodynamic properties of such model are quite unusual, as we know from earlier work \cite{Fisher,Suzuki91}. In particular, we find that the standard spin-spin correlation function of this model demonstrates
persistent incommensurate oscillations Eq.~\eqref{GCLY} on the line of the Lee-Yang zeros with the wavevector which is twice of $q_m$ given by Eq.~\eqref{qm}. This is a sign of a true  long-range order in the oscillating phase  $h \leq h_{c,m}(K)$.

%
%
%%%%%%%%%%%%%%%%%%%%%%%%%%%%%%%%%%%%%%%%%%%%%%%%%%%%%%%%%%%%%%%%%%%%%%%%%%%%%%
\section{Conclusions}\label{Concl}
%%%%%%%%%%%%%%%%%%%%%%%%%%%%%%%%%%%%%%%%%%%%%%%%%%%%%%%%%%%%%%%%%%%%%%%%%%%%%%
%
%
%%%%%%%%%%%%%%%%%%%%%%%%%%%%%%%%%%%%%%%%%%%%%%%%%%%%%%%%%%%%%%%%%%%%%%%%%%%%%%%%%%%%%%%%%%%%%%%%%%%%%
%%%%%%%%%%%%%%%%%%%%%%%%%%%%%%%%%%%%%%%%%%%%%%%%%%%%%%%%%%%%%%%%%%%%%%%%%%%%%%%%%%%%%%%%%%%%%%%%%%%%%
%%%%%%%%%%%%%%%%%%%%%%%%%%%%%%%%%%%%%%%%%%%%%%%%%%%%%%%%%%%%%%%%%%%%%%%%%%%%%%%%%%%%%%%%%%%%%%%%%%%%%
%%%%%%%%%%%%%%%%%%%%%%%%%%%%%%%%%%%%%%%%%%%%%%%%%%%%%%%%%%%%%%%%%%%%%%%%%%%%%%%%%%%%%%%%%%%%%%%%%%%%%

We presented new results for the classical antiferromagnetic Ising chain in a magnetic field.
We show that the model possesses an infinite sequence of phase transitions on the disorder lines.
The phase transition is signalled by a change of asymptotic behavior of the nonlocal correlation
functions of rarefied $m$-periodic strings ($\forall~m =2l+1,~l \in \mathbb{Z}^+$) when their
monotonous decay becomes modulated by incommensurate oscillations.
We proposed a duality transformation which maps the Ising chain onto the $m$-leg Ising tube with nearest-neighbor
couplings along the legs and the plaquette four-spin interactions of adjacent legs. Then the $m$-string correlation
functions of the Ising chain are mapped onto the spin-spin correlation functions along the legs of the $m$-leg tube.
The spin-tube dual Hamiltonians we found are a very interesting example of solvable Ising models with multi-spin interactions. \cite{Turban2016}
We also find that the spin-spin correlation function calculated for this model in the complex magnetic field demonstrates
persistent incommensurate oscillations on the line of the Lee-Yang zeros with the wavevector which is twice of that of the decaying oscillations.

We trace the origin of these cascades of  phase transitions to
the lines of the Lee-Yang zeros of the Ising chain in $m$-periodic complex magnetic field. 
Such interpretation of the disorder line transitions relates the Lee-Yang zeros in complex fields to the observable 
(and potentially measurable) quantities.
We believe that the model possesses even more
disorder lines if more general types of rarefied strings and their mutual correlations are considered. A very promising direction for future work would be generalization of the present results on higher dimensions and/or quantum spin models.

%%%%%%%%%%%%%%%%%%%%%%%%%%%%%%%%%%%%%%%%%%%%%%%%%%%%%%%%%%%%%%%%%%%%%%%%%%%%%%
%%%%%%%%%%%%%%%%%%%%%%%%%%%%%%%%%%%%%%%%%%%%%%%%%%%%%%%%%%%%%%%%%%%%%%%%%%%%%%
%------------------------------------------------------------------------------
\begin{acknowledgments}
We acknowledge financial support from the Laurentian University Research Fund
(LURF) (G.Y.C.) and from the Ministry of Education and
Science of the Russian Federation (state assignment grant No. 3.5710.2017/BCh)
(P.N.T.). G.Y.C. thanks the Centre for Physics of Materials at McGill University
for hospitality.
\end{acknowledgments}
%%%%%%%%%%%%%%%%%%%%%%%%%%%%%%%%%%%%%%%%%%%%%%%%%%%%%%%%%%%%%%%%%%%%%%%%%%%%%%
%%%%%%%%%%%%%%%%%%%%%%%%%%%%%%%%%%%%%%%%%%%%%%%%%%%%%%%%%%%%%%%%%%%%%%%%%%%%%%

%%%%%%%%%%%%%%%%%%%%%%%%%%%%%%%%%%%%%%%%%%%%%%%%%%%%%%%%%%%%%%%%%%%%%%%%%%%%%%
%%%%%%%%%%%%%%%%%%%%%%%%%%%%%%%%%%%%%%%%%%%%%%%%%%%%%%%%%%%%%%%%%%%%%%%%%%%%%%
\begin {appendix}
\section{Transfer matrices and eigenvalues}\label{AppA}

It is convenient \cite{Lancaster69} to write the transfer matrix \eqref{U} via
two orthogonal idempotent operators (projectors) $\hat{\mathcal{P}}_\pm$ as
\begin{equation}
\label{Upm}
  \hat{U} =\lambda_+  \hat{\mathcal{P}}_+  + \lambda_- \hat{\mathcal{P}}_- ~,
\end{equation}
where
\begin{widetext}
\begin{equation}
\label{Epm}
  \hat{\mathcal{P}}_\pm \equiv \pm \frac{\hat U -\lambda_\mp \hat{\mathbb{1}} }{\lambda_+ - \lambda_-}
  = \frac12 \Big\{ \hat{\mathbb{1}}  \pm \frac{ \hat{\sigma}_z \sinh h +e^{2K}( \hat{\sigma}_x \cosh h+i \hat{\sigma}_y \sinh h) }{R(h)} \Big\}
\end{equation}
\end{widetext}
Then $\forall~ m \in \mathbb{N}$:
\begin{equation}
\label{Um}
  \hat{U}^m =\lambda_+^m  \hat{\mathcal{P}}_+  + \lambda_-^m \hat{\mathcal{P}}_- ~.
\end{equation}
Using Eqs.~(\ref{U},\ref{Lampm},\ref{Epm},\ref{Um}) we obtain  for
\begin{widetext}
\begin{equation}
\label{Vm}
 \hat V_m   \equiv  \hat{\sigma}_z \hat{U}^m =
  \frac{\lambda_+^m - \lambda_-^m}{2 R} \Big(\sinh h \big[ \hat{\mathbb{1}}+  \hat{\sigma}_x e^{2K} \big] + i \hat{\sigma}_y e^{2K} \cosh h \Big) +
  \frac12 \big( \lambda_+^m + \lambda_-^m \big) \hat{\sigma}_z~.
\end{equation}
\end{widetext}
The eigenvalues of this matrix are readily found:
\begin{equation}
\label{mumpm}
\mu_m^\pm = \frac12 \Big( \mathrm{Tr} \hat V_m \pm \sqrt{ (\mathrm{Tr}\hat V_m)^2 +4 (\lambda_+ \lambda_-)^m } \Big)~,
\end{equation}
with
\begin{equation}
\label{TrVm}
  \mathrm{Tr} \hat V_m =  \frac{\sinh h}{ R} \big( \lambda_+^m - \lambda_-^m \big)~.
\end{equation}
Since $\lambda_+ \lambda_- = -(e^{4K}-1) $ is negative in the range of parameters under
consideration, the complex eigenvalues (and, consequently, the disorder lines of phase transitions) are possible only for odd $m$:
\begin{equation}
\label{mS}
  m=2l+1,~~ l \in \mathbb{Z}^+~.
\end{equation}

Let us introduce the hyperbolic parametrization as
\begin{widetext}
\begin{equation}
\label{DeltaZeta}
  \zeta \equiv  \mathrm{arcsinh} \Big( \frac{\cosh h}{\Delta} \Big) = \log \frac{\cosh h + \sqrt{\cosh^2 h + \Delta^2}}{\Delta }
  , ~~\mathrm{where} ~~ \Delta \equiv \sqrt{e^{4K}-1}~.
\end{equation}
\end{widetext}
Then the eigenvalues \eqref{mumpm} of the matrix $\hat V_m$ \eqref{Vm} can be written as
\begin{equation}
\label{mueps}
\mu_m^\pm = \Delta^m \Big( \gamma_m \pm \sqrt{\gamma_m^2 -1} \Big)~,
\end{equation}
where
\begin{equation}
\label{eps}
   \gamma_m  \equiv \frac{\sinh h \, \cosh m \zeta}{\Delta \cosh \zeta} =
   \tanh h \, \tanh \zeta \, \cosh m \zeta~.
\end{equation}
For each given $m$ the critical line of the phase transition (disorder line) defined by Eq.~\eqref{Tcm} and shown in  Fig.~\ref{CritLines} corresponds to the condition $\gamma_m=1$.
The condition $\gamma_m>1$ corresponds to the disordered phase, while $\gamma_m<1$ in the oscillating phase.
From the above equations we readily obtain:
\begin{equation}
\mu_m^\pm = \Delta^m
\left\{
\begin{array}{lr}
\exp (\pm  \mathrm{arccosh}  \gamma_m ),
& \gamma_m>1~; \\
\exp (\pm i   \arccos  \gamma_m ),
& \gamma_m<1~.
\end{array}
\right. \label{Mupm}
\end{equation}
For $m=1$ the equation for the disorder line \eqref{Tcm} reduces to the Stephenson result \eqref{hc}. \cite{Stephenson70}
One can also find the explicit transcendental equation for the critical line at $m=3$:
\begin{widetext}
\begin{equation}
\label{hc3}
 \sinh h_{c,3}(K)= \frac{\Delta}{2^{1/3}} \Bigg[ \Bigg( \sqrt{1+ \frac{1}{432} \frac{(\Delta^2+4)^3}{\Delta^6}}  +  1 \Bigg)^{\frac13}
        -  \Bigg( \sqrt{1+ \frac{1}{432} \frac{(\Delta^2+4)^3}{\Delta^6}}  - 1 \Bigg)^{\frac13} \Bigg]~.
\end{equation}
\end{widetext}
The critical lines  for $m>3$ are found by numerical solution of Eqs.~\eqref{Tcm} and \eqref{DeltaZeta}.

%
%%%%%%%%%%%%%%%%%%%%%%%%%%%%%%%%%%%%%%%%%%%%%%%%%%%%%%%%%%%%%%%%%%%%%%%%%%%%%%
%%%%%%%%%%%%%%%%%%%%%%%%%%%%%%%%%%%%%%%%%%%%%%%%%%%%%%%%%%%%%%%%%%%%%%%%%%%%%%
\section{Correlation functions}\label{AppB}

To calculate correlation functions of the rarefied strings  \eqref{GnmUV} we will need powers of the operator
$\hat V_m$.  It is most conveniently done by using two orthogonal projectors $\hat{\mathcal{P}}_m^{\pm}$ as
\begin{equation}
\label{Vpm}
  \hat{V}_m =\mu_m^+  \hat{\mathcal{P}}_m^+  + \mu_m^-  \hat{\mathcal{P}}_m^- ~,
\end{equation}
where
\begin{equation}
\label{Ppm}
 \hat{\mathcal{P}}_m^{\pm} \equiv \pm \frac{\hat{V}_m -\mu_m^{\mp} \hat{\mathbb{1}} }{\mu_m^+ - \mu_m^-}~.
\end{equation}
Then $\forall~ n \in \mathbb{N}$:
\begin{equation}
\label{Vmn}
  (\hat{V}_m)^n =(\mu_m^+)^n \hat{\mathcal{P}}_m^+  + (\mu_m^-)^n \hat{\mathcal{P}}_m^- ~.
\end{equation}
For odd $m$ the explicit formula for the projectors reads:
\begin{equation}
\label{PpmOdd}
 \hat{\mathcal{P}}_m^{\pm} =\frac12
 \Big\{ \hat{\mathbb{1}}  \pm \frac{ \hat{\sigma}_z \sinh m \zeta +\gamma_m e^{2K}
 ( \hat{\sigma}_x +i \hat{\sigma}_y /\tanh h) }{\sqrt{\gamma_m^2 -1}} \Big\}~.
\end{equation}
With all these formulas it is straightforward to get the correlation function of the rarefied strings  \eqref{GnmUV}
in the thermodynamic limit:
\begin{equation}
\label{GnmLim}
   G(n|m)~ \xrightarrow{N \to \infty}~ g_m^+ \bigg(  \frac{\mu_m^+}{\lambda_+^m}  \bigg)^n
   + g_m^- \bigg(  \frac{\mu_m^-}{\lambda_+^m}  \bigg)^n~,
\end{equation}
where
\begin{equation}
\label{gpm}
   g_m^\pm \equiv \mathrm{Tr} \big\{ \hat{\mathcal{P}}_+  \hat{\mathcal{P}}_m^\pm \big\}
   =\frac12
 \bigg\{ 1 \pm \frac{ \gamma_m \tanh m \zeta}{\sqrt{\gamma_m^2 -1}} \bigg\}
    ~.
\end{equation}
In the disordered phase where  $\gamma_m>1$, it decays exponentially and monotonously
\begin{widetext}
\begin{equation}
\label{GnmAbove}
   G(n|m) =  e^{-\zeta mn} \Big[ \cosh mn(\zeta -\kappa_m) + \sinh mn(\zeta -\kappa_m)
   \frac{\tanh m \zeta}{\tanh m(\zeta -\kappa_m)} \Big]
   \sim e^{- \kappa_m r}
\end{equation}
\end{widetext}
with the physical distance $r=nm \gg 1 $ between the right ends of the rarefied strings. The inverse correlation length
\begin{equation}
\label{kappaAb}
  \kappa_m = \frac1m \log  \frac{\lambda_+^m}{\mu_m^+}=  \zeta - \frac1m  \mathrm{arccosh}  \gamma_m~.
\end{equation}
In the oscillating phase when $\gamma_m<1$,  the correlation function  behaves as
\begin{widetext}
\begin{equation}
\label{GnmBelow}
  G(n|m) = e^{-\kappa_m mn} \Big[ \cos(mnq_m) + \sin (mn q_m)
   \frac{\tanh m \zeta}{\tan m q_m} \Big]
  \sim e^{- \kappa_m r} \cos(q_m r +\delta_m)
\end{equation}
\end{widetext}
at $r \gg 1$. The inverse correlation length in this case is
\begin{equation}
\label{kappaBe}
  \kappa_m = \frac1m \log  \frac{\lambda_+^m}{|\mu_m^+|}=  \zeta ~,
\end{equation}
and the incommensurate wave vector of the oscillations
\begin{equation}
\label{qm}
  q_m = \frac1m \arg \mu_m^+=  \frac1m \arccos  \gamma_m~.
\end{equation}
At the critical point $\gamma_m=1$
\begin{equation}
\label{GnmTc}
  G(n|m) = e^{-\kappa_m mn} \Big[1+ n \tanh m \zeta \Big]
\end{equation}
For $m=1$ the above equations recover the result of Stephenson  \eqref{q}. \cite{Stephenson70}
Examples of the inverse correlation lengths and wave vectors of oscillations numerically calculated from the above equations
are plotted in Fig.~\ref{kappaFig} and Fig.~\ref{qFig}.

%
%%%%%%%%%%%%%%%%%%%%%%%%%%%%%%%%%%%%%%%%%%%%%%%%%%%%%%%%%%%%%%%%%%%%%%%%%%%%%%
%%%%%%%%%%%%%%%%%%%%%%%%%%%%%%%%%%%%%%%%%%%%%%%%%%%%%%%%%%%%%%%%%%%%%%%%%%%%%%
\section{ Ising model in $m$-periodic complex field}\label{AppC}

We will be interested in the special case of the $m$-periodic magnetic field $\mathfrak{h}_m(k)$ defined by \eqref{Hm}
when  $\mathfrak{h}= h +i \pi/2$, i.e., the field on the complex plane $\mathfrak{h}$ varies along the line parallel to the real
axis and passing through the locus of the Lee-Yang zeros, as shown in  Fig.~\ref{H_plane}. For convenience we will use the upper case
for this special field, i.e.
\begin{equation}
\mathfrak{H}_m (k)=
\left\{
\begin{array}{lr}
h+ i \frac{\pi}{2},
& k=m l~ \& ~l\in \mathbb{N} ; \\
h
& \mathrm{elswhere}~.
\end{array}
\right. \label{HCs}
\end{equation}
We are interested in calculating the correlation function of spins residing on the sites where the complex field is applied:
\begin{equation}
\label{GCnDef}
  \mathcal{G}(nm,\mathfrak{H}_m ) \equiv \langle s(ml)s(ml+mn) \rangle~.
\end{equation}
Using Eqs.~(\ref{VmCompl},\ref{LYz}) we get
\begin{widetext}
\begin{equation}
\label{GCn}
  \mathcal{G}(nm,\mathfrak{H}_m )= \mathrm{Tr} \Big\{ \hat{\sigma}_z
  \Big( \hat{\mathcal{U}}_m(\mathfrak{H}_m) \Big)^n \hat{\sigma}_z \Big( \hat{\mathcal{U}}_m(\mathfrak{H}_m) \Big)^{N_L-n}
   \Big\}/Z(\mathfrak{H}_m)=
   \frac{ \mathrm{Tr} \big\{\hat{\sigma}_z \big( \hat{V}_m(h)\big)^n \hat{\sigma}_z \big( \hat{V}_m(h)\big)^{N_L-n} \big\}}{(\mu_m^+)^{N_L}+(\mu_m^-)^{N_L}}
  ~.
\end{equation}
\end{widetext}
The ends of the Lee-Yang line $h = \pm h_{c,m}(K)$ ($\gamma_m=1$) are the critical points of the model, and in the same time they are
the branch points of  $\mu_m^\pm(h)$, cf. Eq.~\eqref{mueps} and Fig.~\ref{H_plane}. Between the critical points $\pm h_{c,m}(K)$ along the Lee-Yang line the partition function oscillates rapidly and does not have a rigorous thermodynamic limit. This problem is known from earlier work \cite{Fisher,Suzuki91}. We handle it by cutting the complex plane $\mathfrak{h}$ between the branch point along the Lee-Yang line (blue bold line in Fig.~\ref{H_plane}).
Addition of an infinitesimal imaginary part $ h \rightarrow h+i \epsilon$ ($\epsilon \to 0^+$) results in
\begin{equation}
\label{muCond}
  |\mu_m^+(h+i \epsilon)| > |\mu_m^-(h+i \epsilon)|
\end{equation}
and allows us to take the thermodynamic limit and calculate parameters in the oscillating phase when $|h| < h_{c,m}(K)$ by reaching the cut in the complex plane $\mathfrak{h}$ from the top. This yields
\begin{widetext}
\begin{equation}
\label{GnmC}
  \mathcal{ G}(nm,\mathfrak{H}_m  )~ \xrightarrow{N \to \infty}~
  \mathrm{Tr}\big\{ \big( \hat{\sigma}_z  \hat{\mathcal{P}}_m^+ \big)^2 \big\}+
   \mathrm{Tr}\big\{ \hat{\sigma}_z  \hat{\mathcal{P}}_m^+  \hat{\sigma}_z  \hat{\mathcal{P}}_m^-\big\}
   \bigg(  \frac{\mu_m^-}{\mu_m^+}  \bigg)^n  ~.
\end{equation}
\end{widetext}
One can show after some algebra that
\begin{equation}
\label{Tr1}
  \mathrm{Tr}\big\{ \big( \hat{\sigma}_z  \hat{\mathcal{P}}_m^+ \big)^2 \big\}
  =M(\mathfrak{H}_m)^2~,
\end{equation}
where we introduced the magnetization on the sites $k=ml$ where the field $h+i\pi/2$ is applied:
\begin{equation}
\label{Mc}
M(\mathfrak{H}_m) \equiv  \langle s(ml)\rangle =\frac{\sinh m \zeta}{\sqrt{\gamma_m^2-1}}~.
\end{equation}
The second matrix term in \eqref{GnmC} can be also worked out into a simple form
\begin{equation}
\label{Tr2}
  \mathrm{Tr}\big\{ \hat{\sigma}_z  \hat{\mathcal{P}}_m^+  \hat{\sigma}_z  \hat{\mathcal{P}}_m^-\big\}
  =1- M(\mathfrak{H}_m)^2~,
\end{equation}
yielding a nice result for the correlation function:
\begin{equation}
\label{GnmSimple}
  \mathcal{ G}(nm,\mathfrak{H}_m ) = M(\mathfrak{H}_m)^2 + \Big[1- M(\mathfrak{H}_m)^2 \Big]
   \bigg(  \frac{\mu_m^-}{\mu_m^+}  \bigg)^n  ~.
\end{equation}
Subtracting  a trivial constant term due to the field-induced magnetization, we readily obtain exponentially
dying off correlations when $|h| > h_{c,m}(K)$:
\begin{equation}
\label{GCabove}
  \mathcal{ G}_R(r,\mathfrak{H}_m ) \equiv \mathcal{ G}(r,\mathfrak{H}_m )- M(\mathfrak{H}_m)^2
   =  \big[1- M(\mathfrak{H}_m)^2 \big] e^{-\tilde{\kappa}_m r}~,
\end{equation}
where  $r=nm$ and the inverse correlation length
\begin{equation}
\label{kappaTilde}
  \tilde{\kappa}_m =  \frac2m  \mathrm{arccosh}  \gamma_m= \frac2m \log \big( \gamma_m \pm \sqrt{\gamma_m^2 -1} \big)~.
\end{equation}
Along the Lee-Yang line (branch cut) when $|h| < h_{c,m}(K)$ the correlations do not decay, but oscillate steadily as
\begin{equation}
\label{GCLY}
  \mathcal{ G}_R(r,\mathfrak{H}_m )
   =  \big[1- M(\mathfrak{H}_m)^2 \big] e^{-i2 q_m r}~,
\end{equation}
where the wavevector $q_m$ is given by Eq.~\eqref{qm}. Such behavior of the correlation function $\mathcal{ G}_R(r,\mathfrak{H}_m )$
can be interpreted as a sign of the local long-range spin order in the model with complex field in the oscillating phase. The critical behavior
of this model near the critical point $|h| = h_{c,m}(K)$ ($\gamma_m=1$) is quite unusual: although the correlation length diverges
($\tilde{\kappa}_m \to 0$ in Eq.~\eqref{kappaTilde}) as one would expect, so does the magnetization $M(\mathfrak{H}_m)$. Moreover, the
latter stays purely imaginary along the Lee-Yang line. These findings are in agreement with earlier results by Fisher and Suzuki for
the homogeneous complex field ($m=1$). \cite{Fisher,Suzuki91}

\end{appendix}
%%%%%%%%%%%%%%%%%%%%%%%%%%%%%%%%%%%%%%%%%%%%%%%%%%%%%%%%%%%%%%%%%%%%%%%%%%%%%%
%%%%%%%%%%%%%%%%%%%%%%%%%%%%%%%%%%%%%%%%%%%%%%%%%%%%%%%%%%%%%%%%%%%%%%%%%%%%%%

%%%%%%%%%%%%%%%%%%%%%%%%%%%%%%%%%%%%%%%%%%%%%%%%%%%%%%%%%%%%%%%%%%%%%%%%%%%%%%
%%%%%%%%%%%%%%%%%%%%%%%%%%%%%%%%%%%%%%%%%%%%%%%%%%%%%%%%%%%%%%%%%%%%%%%%%%%%%%
%------------------------------------------------------------------------------

%
%
%%%%%%%%%%%%%%%%%%%%%%%%%%%%%%%%%%%%%%%%%%%%%%%%%%%%%%%%%%%%%%%%%%%%%%%%%%%%%%
%%%%%%%%%%%%%%%%%%%%%%%%%%%%%%%%%%%%%%%%%%%%%%%%%%%%%%%%%%%%%%%%%%%%%%%%%%%%%%

%
%
\end{document}